\documentstyle[prb,epsfig,aps,multicol,cite]{revtex}
\draft
\author{{R. Folk$^{(1)}$, Yu. Holovatch$^{(2)}$, and T. Yavors'kii$^{(3)}$}
\\[1.5ex]
$^{(1)}$Institut f\"ur Theoretische Physik,
Johannes Kepler Universit\"at  \\ Linz, Altenbergerstrasse 69,
A--4040 Linz, Austria. e-mail: folk@tphys.uni-linz.ac.at
\\
$^{(2)}$Institute for Condensed Matter Physics,
National Academy \\ of Sciences of Ukraine,
1~Svientsitskii Str., UA--79011 Lviv, Ukraine and
\\
Ivan Franko National University of Lviv,
12~Drahomanov Str., UA--79005 Lviv, Ukraine.
\\
e-mail: hol@icmp.lviv.ua
\\
$^{(3)}$Ivan Franko National University of Lviv,
\\
12~Drahomanov Str., UA--79005 Lviv, Ukraine.
e-mail: tarasyk@ktf.franko.lviv.ua}
\title{
\Large\bf Pseudo--$\varepsilon$ expansion of six--loop renormalization 
group functions of an anisotropic cubic model}
\date{\today}

\begin{document}

\maketitle

\begin{abstract}
{\small
Six--loop massive scheme renormalization group functions of a
$d=3$--dimensional cubic model
({\em J.~M.~Carmona, A.~Pelissetto, and E.~Vicari, 
Phys. Rev. B {\bf 61}, 15136 (2000)})
are reconsidered by means of the pseudo--$\varepsilon$ expansion.
The marginal order parameter components number
$N_c=2.862 \pm 0.005$ as well as critical exponents of the cubic model are 
obtained. Our estimate $N_c<3$ leads in particular
to the conclusion that all ferromagnetic cubic crystals
with three easy axis should undergo a first order phase transition.
}
\end{abstract}

\section{Introduction}
Progress in the qualitative understanding and the quantitative 
description of critical phenomena to a great extent was achieved by 
the ideas of renormalization group (RG) theory \cite{RG}. Only global 
features of a many--body system such as the range of interparticle 
forces, the space dimensionality $d$ as well as the dimension $N$ and 
the symmetry of an order parameter were suggested to be responsible 
for long--distance and abrupt behaviour of matter in the critical 
region. As a final step the role of the relevant parameters in 
microscopic Hamiltonians of various nature was represented adequately 
by effective Hamiltonians used in field theories.  While already a 
vector field theory with an isotropic rotationally symmetrical order 
parameter allowed unified and correct description of a large spectrum 
of critical phenomena, an extension of theories is of special interest 
since in real substances anisotropies are always present 
\cite{Aharony76}.  For instance, in cubic crystals one expects the 
spin interaction to react to the lattice structure (crystalline
anisotropy), suggesting additional terms in the Hamiltonian,
invariant under the cubic group. The anisotropy breaks rotational
symmetry of the Heisenberg--like ferromagnet and makes the order
parameter to point either along edges or along diagonals of a cube. 
The corresponding field theory is defined by a Landau - Ginzburg - 
Wilson (LGW) Hamiltonian with two $\phi^4$ terms of $O(N)$ {\bf and} 
cubic symmetry and can exhibit a second order phase transition 
characterized either by spherical {\bf or} cubic critical exponents. 
Varying the number of components of the order parameter $N$ a new 
crossover phenomenon between these two scenarios takes place at the 
marginal value $N_c$.

In the framework of RG theory the critical point corresponds to the 
stable fixed point of the RG transformation \cite{RG}.  In a model 
with competing fixed points the study of domains of their attraction 
as well as the crossover phenomenon is a fundamental problem for 
universality comprehension. Apart from the academic interest the 
determination of $N_c$ can lead to decisive conclusions about phase 
transition order in a certain class of cubic crystals. For instance, 
$d=3$ cubic crystals with three easy axis should undergo either a 
second or a weak first--order phase transition provided $N_c$ is 
greater or less than $3$ \cite{Sznajd}.  This argumentation states 
that the existence of the stable fixed point of a field theory is a 
necessary but not a sufficient condition for a model to exhibit a 
second order phase transition. If the parameters of an microscopic 
Hamiltonian are mapped in the plain of the LGW Hamiltonian couplings 
to a point which lies outside the domain of attraction of the stable 
fixed point, the Hamiltonians will flow away to infinitely large 
values of couplings. Such a behaviour might serve as evidence of a weak 
first order phase transition and this is confirmed in some experiments 
(see Ref. \cite{Sznajd} and references therein). If $N_c<3$ for a 
$d=3$ ferromagnet, the new cubic fixed point governing the critical 
regime appears to be inaccessible from the initial values of couplings 
which correspond to the ferromagnetic ordering with three easy axis.
It appears that $N_c$ is very close to $N=3$ and the critical 
exponents in both regimes are indistinguishable experimentally. In 
order to calculate the value of $N_c$ within field theory one has to 
treat a complex model of two couplings.  This is different from a 
Heisenberg-like $N$--component ferromagnet with weak quenched 
disorder, where the Harris criterion answers the question about the 
type of critical behaviour \cite{HarrisCriterion}.

The description of the crossover and the precise determination of its 
numerical characteristics has been a challenge for many RG studies of 
the anisotropic cubic model. High orders of perturbation theory were 
obtained for this model in successive approximations either in 
$\varepsilon$--expansion with dimensional regularization in the 
minimal subtraction (MS) scheme \cite{MS} or within the massive $d=3$ 
scheme \cite{Parisi}. The expressions are available now in the 
five--loop \cite{Kleinert95a} and in the six--loop approximations 
\cite{Carmona99} respectively.  However, divergent properties of the 
series did not allow their straightforward analysis and called for the 
application of various resummation procedures. For instance, $N_c$ 
calculated in five--loop $\varepsilon$--expansion \cite{Kleinert95a} 
yielded depending on the resummation procedure: $N_c=2.958$ 
\cite{Kleinert95a}, $N_c=2.855$ \cite{Shalaev97} and $N_c=2.87(5)$ 
\cite{Carmona99}. Alternative approaches on the basis of the 
$\varepsilon$--expansion lead to $N_c=2.97(6)$ \cite{Carmona99} and to 
$N_c=2.950$ \cite{Holovatch99a}.  On the other hand the massive $d=3$ 
scheme RG functions extended for arbitrary $N$ to four loops 
\cite{Varnashev99}, yielded $N_c=2.89(2)$ \cite{Varnashev99} (see 
\cite{Carmona99} for a recent extended review of theoretical 
determination of $N_c$).  These results suggest that the most reliable 
theoretical estimate is $N_c<3$.

However, recent MC simulations \cite{Caselle98} questioned the values 
for $N_c$ obtained so far. There, considering the finite size 
corrections of a cubic invariant perturbation term at the critical 
$O(N)$--symmetric point, the eigenvalues $\omega_i$ of the stability 
matrix were extracted.  From the estimate $\omega_2=0.0007(29)$ a 
value of $N_c=3$ was concluded. This disagreement as well as the 
crucial influence of the value of  $N_c$ on the order of the phase 
transition makes an implementation of an alternative method for 
calculation of $N_c$ to be hardly overrated.  Recently the massive 
$d=3$ scheme RG functions of the cubic model were extended to 
five--loop order \cite{Pakhnin99} and very recently the six--loop 
series \cite{Carmona99} were obtained. The traditional analysis of 
these series, including an information on large order behaviour of the 
RG functions \cite{Carmona99}, yielded $N_c=2.89(4)$.  However let us 
note that the most accurate estimates of the critical exponents of a 
$d=3$ $O(N)$--symmetric $\phi^4$ model in a massive field theoretical 
RG scheme are based on a pseudo--$\varepsilon$ expansion technique 
\cite{LeGuillou80,Guida98}. Up to our knowledge the last method has never been 
applied to the cubic model \cite{VonFerber}.  Therefore the main aim 
of the present paper is to apply the pseudo--$\varepsilon$ expansion 
to the up--to--date most precise massive scheme RG function 
\cite{Carmona99} of the cubic model.

The set-up of the article is as follows. After a brief consideration 
of the model and renormalization procedure, we present the  
pseudo--$\varepsilon$ expansion for $N_c$ and discuss its properties. 
Applying Pad\'e-- and Pad\'e--Borel analysis we obtain precise 
estimates of $N_c$ and compare the result with the corresponding 
$\varepsilon$--expansion.  Finally we evaluate the critical exponents 
of a $d=3$ cubic system belonging to the new universality class for 
different values of $N>N_c$ and discuss the weakly diluted Ising model 
case $N=0$.

\section{Pseudo-$\varepsilon$ series and numerical results}
We start from a $d=3$ effective LGW Hamiltonian with two couplings at 
terms of spherical and cubic symmetry:
\begin{equation}
\label{1}
{\cal H}(\varphi)=\int {\rm d}^3R \Big\{ {1\over 2} 
\sum_{\alpha=1}^{N} \left[|\nabla {\varphi}_\alpha|^2+ m_0^2 
{\varphi}_\alpha^2\right] + {u_{0}
\over4!} \left(\sum_{\alpha=1}^{N}{\varphi}_\alpha^2 \right)^2 +
{v_{0}
\over 4!} \sum_{\alpha=1}^{N}{\varphi}_\alpha^4 \Big\},
\end{equation}
where $\varphi_\alpha(R)$ are components of a bare
$N$--component vector field;
$u_{0} > 0$, $v_{0}$ are bare couplings, $m_0^2$ is a squared bare
mass being a linear function of temperature. The vicinity of a
critical point corresponds to a long-distance behaviour of the model
(\ref{1}), while ultraviolet divergences of the theory are dealt
with by means of an appropriate renormalization procedure
\cite{RG}. In particular, the renormalization of the bare couplings 
leads to $\beta_u(u, v)$ and $\beta_v(u, v)$ -- the so-called 
$\beta$--functions; a renormalization of the bare field and
square--field insertion produces $\gamma_{\phi}(u, v)$ and
$\bar{\gamma}_{\phi^2}(u, v)$ -- the so-called $\gamma$--functions.
All these functions depend on renormalized couplings $u$ and $v$
\cite{dependencefree}. The critical behaviour of the model is 
determined by the infrared stable fixed point $u^*, v^*$. It is given 
by the condition that both $\beta$--functions are zero and all the 
real parts of the stability matrix eigenvalues are positive. The pair 
correlation function critical exponent $\eta$ and the correlation 
length critical exponent $\nu$ are obtained via the relations $\eta = 
\gamma_{\phi}(u^*, v^*), 1/\nu = 2-\bar{\gamma}_{\phi^2}(u^*, 
v^*)-\gamma_{\phi}(u^*, v^*)$.  The correction-to-scaling exponent 
$\omega$ is given by the largest stability matrix eigenvalue in the 
stable fixed point.

In the present study we reconsider RG functions of the model 
(\ref{1}) as they are obtained within massive fixed $d=3$ scheme 
\cite{normalization}:
\begin{eqnarray}
-{\frac {\beta_u(u,v)}{u}} &=&
1- u - {\frac {2}{3}} v
+{\frac {4}{27}}{\frac {\left (190+41 N\right )}{\left (8+N\right 
)^{2}}}{u}^{2} +{\frac {400}{81}}{\frac {uv}{8+N}} +{\frac 
{92}{729}}{v}^{2} + \beta_u^{(3LA)} + \ldots + \beta_u^{(6LA)},
\label{beta_u}
\\
-{\frac {\beta_v(u,v)}{v}} &=&
1-12 {\frac {u}{8+N}}-v
+{\frac {4}{27}}{\frac {\left (370+23 N \right )}{\left (8+N\right 
)^{2}}}{u}^{2} +{\frac {832}{81}}{\frac {uv}{8+N}} +{\frac 
{308}{729}}{v}^{2} + \beta_v^{(3LA)} + \ldots + \beta_v^{(6LA)},
\label{beta_v}
\\
\gamma_{\phi}(u,v) &=&
{\frac {8}{27}}{\frac {\left (2+N\right )}{\left (8+N\right ) 
^{2}}}{u}^{2} +{\frac {16}{81}}{\frac {uv}{8+N}} +{\frac 
{8}{729}}{v}^{2} + \gamma_{\phi}^{(3LA)} + \ldots + 
\gamma_{\phi}^{(6LA)}, \label{gamma_phi}
\\
\bar{\gamma}_{\phi^2}(u,v) &=&
{\frac {\left (2+N\right )u}{8+N}}
+{\frac {1}{3}}v
-2{\frac {\left (2+N\right )}{\left (8+N\right )^2}}{u}^{2}
-{\frac {4 uv}{3 (8+N)}}
-{\frac {2}{27}}{v}^{2}
+ \bar{\gamma}_{\phi^2}^{(3LA)} + \ldots + 
\bar{\gamma}_{\phi^2}^{(6LA)}, \label{gamma_phi2} \end{eqnarray}
where $\beta_u^{(3LA)} \dots \bar{\gamma}_{\phi^2}^{(6LA)}$
denote the three--loop contributions obtained in Ref. \cite{3loops},
the four--loop, the recent five--loop and
the very recent six--loop contributions obtained in Refs. 
\cite{Varnashev99}, \cite{Pakhnin99} and \cite{Carmona99} 
respectively. Furthermore, the large-order behaviour was established 
for the RG functions (\ref{beta_u})--(\ref{gamma_phi2}) which allowed 
to apply the refine resummation procedure based on Borel 
transformation with conformal mapping \cite{Carmona99}. In this way a 
convergent sequence of approximations for $N_c$ as well as critical 
exponents within the cubic universality class were obtained 
\cite{Carmona99}.

One possible way of analysis of the massive RG functions
(\ref{beta_u})--(\ref{gamma_phi2})
consists in solution of a system of equations for the (resummed)
$\beta$--functions (\ref{beta_u}), (\ref{beta_v})
\begin{eqnarray}
\beta_u(u^*,v^*) &=& 0,
\nonumber
\\
\beta_v(u^*,v^*) &=& 0
\end{eqnarray}
to get numerical values of a stable fixed point coordinates $u^*,v^*$.
Then these numerical values are substituted into (resummed) series for
$\gamma$--functions (\ref{gamma_phi}), (\ref{gamma_phi2}) which lead 
to the numerical values for critical exponents.  As a result the final 
errors for the critical exponents are the sum of the errors of the 
series for exponents and of the errors coming from $u^*,v^*$. To avoid 
such errors accumulation it is standard now in the analysis of field
theories with one coupling \cite{LeGuillou80,Guida98}
to use a pseudo--$\varepsilon$ expansion \cite{Nickel}. Here, we will 
apply the pseudo--$\varepsilon$ expansion for a cubic model (\ref{1}) 
\cite{VonFerber}. The procedure is defined in the following way. Let 
us introduce the functions:
\begin{eqnarray}
\beta_u(u,v,\tau) &=& -u(
\tau- u - {\frac {2}{3}} v + \ldots),
\nonumber
\\
\beta_v(u,v,\tau) &=& -v(
\tau-12 {\frac {u}{8+N}}-v + \ldots)
\end{eqnarray}
where the ``pseudo--$\varepsilon$'' auxiliary parameter $\tau$ has 
been introduced into the $\beta$--functions (\ref{beta_u}), 
(\ref{beta_v}) instead of the zeroth order term. Obviously, 
$\beta_u(u,v)\equiv\beta_u(u,v,\tau=1)$, 
$\beta_v(u,v)\equiv\beta_v(u,v,\tau=1)$. Then a fixed point 
coordinates are obtained as series in $\tau$. The series for the 
stable fixed point coordinates  $u^*(\tau)$, $v^*(\tau)$ are then 
substituted into series (\ref{gamma_phi}), (\ref{gamma_phi2}) for 
$\gamma$--functions leading  to the pseudo--$\varepsilon$ expansion 
for critical exponents. In the resulting series the expansion 
parameter $\tau$ collects contributions from the loop integrals of 
the same order coming from both the series of $\beta$-- and 
$\gamma$--functions. Finally, one puts $\tau=1$. In such a way one 
gets a self-consistent perturbation theory and avoids cumulation of 
errors originating from different steps of calculation.

With the above described method we obtain the marginal value $N_c$ and 
the critical exponents in the cubic universality class for the model 
(\ref{1}). The series for $N_c$ reads:

\begin{equation}\label{nc_tau}
N_c = 4 - 4/3\tau + 0.29042005\tau^2 -
                    0.18967704\tau^3 +
                    0.19951035\tau^4 -
                    0.22465150\tau^5.
\end{equation}
One notes that at least up to the presented number of loops the series 
does not behave like an asymptotic one with factorial growth of 
coefficients.  This can be seen by considering a Pad\'e--table 
(\ref{nc_pd}) for $N_c$ series (\ref{nc_tau}):
\begin{equation}\label{nc_pd}
\left [\begin {array}{cccccc}  4& 3& 2.9158& 2.8411& ^{2.8922}&2.8298\\
\noalign{\medskip} 2.6667& 2.9051& ^{2.0643}& 2.8711& 2.8638&o\\
\noalign{\medskip} 2.9571& 2.8423& 2.8616& 2.8616&o&o\\
\noalign{\medskip} 2.7674& 2.8646& 2.8616&o&o&o\\
\noalign{\medskip} 2.9669& 2.8613&o&o&o&o\\
\noalign{\medskip} 2.7423&o&o&o&o&o\end {array}
\right ].
\end{equation}
Here, a result of an approximant $[M/N]$ is
represented as an element of a matrix with usual notation. The
approximants $[0/4]$ and $[1/2]$ have poles at values of $\tau$ of the 
order 1 (at points $\tau_1=3.7$ and $\tau_2=1.1$ respectively) and 
thus the estimates of $N_c$ on their basis are considered as 
unreliable (they are noted in the table by small numbers). The values 
in the first column of the table are merely the sums of the 
corresponding number of terms in the expansion (\ref{nc_tau}) and do 
not diverge. However, the most prominent property of the table is the 
perfect convergence of the values within main diagonals. In 
particular, the six--loop result of the $[3/2]$ and the $[2/3]$ 
approximants and the five--loop result of $[2/2]$ approximant coincide 
within the 4th digit and lead to an estimate $N_c=2.8616$.  Though the 
next order terms in (\ref{nc_tau}) could spoil such convergence, it is 
worth to compare the pseudo--$\varepsilon$ expansion (\ref{nc_tau}) 
for $N_c$  with corresponding $\varepsilon$--expansion series which is 
one loop order shorter \cite{Kleinert95a}:
\begin{equation}\label{nc_epsilon}
N_c=4- 2\varepsilon+
 2.58847559\varepsilon^{2}-
 5.87431189\varepsilon^{3}+
16.82703902\varepsilon^{4}.
\end{equation}
The obvious worse convergence properties of the series 
(\ref{nc_epsilon}) lead to a corresponding bad convergence of the 
values in the Pad\'e--table (obtained in Ref. \cite{Kleinert95a}). In 
particular, mere summation of several first terms leads now to 
diverging result:
\begin{equation}
\left [\begin {array}{ccccc}
4& 2.667&^{3.627}& 1.952&^{-5.772}\\
\noalign{\medskip} 2& 3.128& 2.893& 2.972&o\\
\noalign{\medskip} 4.589& 2.792& 2.958&o&o\\
\noalign{\medskip}- 1.286& 3.068&o&o&o\\
\noalign{\medskip} 15.540&o&o&o&o\end {array}
\right ].
\end{equation}
Here, the approximants $[0/2]$, $[0/4]$ have poles close to $\tau=1$ 
(at $\tau_1=2.3$, $\tau_2=0.9$) respectively and thus are unreliable.

In order to take into consideration possible factorial divergence of 
the pseudo--$\varepsilon$ expansion (\ref{nc_tau}), as a next step we 
apply to the series (\ref{nc_tau}) the Pad\'e--Borel resummation 
procedure. The Pad\'e--Borel resummation of the initial sum 
$N_c(\tau)$ consists of the following steps:  i) construction of 
Borel--Leroy image of  $N_c(\tau)$; ii) its extrapolation by a 
rational Pad\'e--approximant $[M/N](\tau t)$ and iii) definition of a 
resumed $N_c^{res}(\tau)$ by the integration $\int_0^{\infty} {\rm d}t 
\exp(-t) t^p [M/N](\tau t)$, where $p$ is an arbitrary parameter 
entering the Borel--Leroy image \cite{Baker}. One possibility to fix 
$p$ is to require fastest convergence of the resulting values, given 
by the diagonal approximant resummation similar to Pad\'e--analysis 
(\ref{nc_pd}).  However the convergence of these values appears to be 
almost independent of $p$.  On the other hand approximants 
possessing poles on the positive real axis are considered as 
unreliable and we can equally well choose $p$ to provide a minimal 
number of such divergent approximants. For instance, for $p\geq 4$ the 
imaginary part is smaller then $10^{-10}$ in the ''bad'' $[1/4]$ and 
$[3/2]$ approximants and therefore can be neglected.  Processing the 
series (\ref{nc_tau}) for $p=4$ as described we obtain the results 
presented in (\ref{nc_pb}). Here, one encounters only one unreliable 
approximant $[1/2]$ which is again denoted with small numbers.  This 
analysis yields $N_c=2.862 \pm 0.005$, where error bar stems from the
maximal deviation between the six and the five--loop results for 
arbitrary $p$ between $0$ and $10$.
\begin{equation}\label{nc_pb}
\left [\begin {array}{cccccc}
                    4& 3.0353& 2.9245& 2.8634& 2.8763& 2.8561\\
\noalign{\medskip} 2.6667& 2.8995& ^{2.7173}& 2.8737& 2.8685&o\\
\noalign{\medskip} 2.9571& 2.8461& 2.8595& 2.8631&o&o\\
\noalign{\medskip} 2.7674& 2.8617& 2.8645&o&o&o\\
\noalign{\medskip} 2.9669& 2.8641&o&o&o&o\\
\noalign{\medskip} 2.7423&o&o&o&o&o\end {array}
\right ].
\end{equation}

It is obvious that other values of interest such as fixed point 
coordinates and critical exponents can also be obtained within the 
pseudo--$\varepsilon$ expansions.  For instance, for different values 
of $N$ we obtain the following expressions for the critical exponents  
$\gamma$ of the susceptibility, $\nu$ of the  correlation length,  and 
$\omega$ of the correction--to--scaling:
\begin{eqnarray}
\gamma_{_{N=3}} &=&
1 + 2/9 \tau + 0.10157666\tau^2+0.03325297\tau^3+
               0.02024452\tau^4+0.00312386\tau^5+
               0.00905558\tau^6,
\nonumber
\\
\gamma_{_{N=4}} &=&
1 + 1/4 \tau + 0.11188272\tau^2+0.03494088\tau^3+
               0.01575673\tau^4-0.00023288\tau^5+
               0.00322125\tau^6,
\nonumber
\\
\gamma_{_{N=5}} &=&
1 + 4/15\tau + 0.11314861\tau^2+0.03107333\tau^3+
               0.00939269\tau^4-0.00376555\tau^5-
               0.00055733\tau^6,
\nonumber
\\
\gamma_{_{N=\infty}}&=&
1 + 1/3\tau+   0.06675812\tau^2+0.00726155\tau^3-
               0.00746706\tau^4-0.00082309\tau^5-
               0.00713623\tau^6,
\label{exp}
\\
\nu_{_{N=3}} &=&
1/2  + 1/9\tau+0.05383664\tau^2+0.01993814\tau^3+
               0.01227945\tau^4+0.00300477\tau^5+
               0.00535272\tau^6,
\nonumber
\\
\nu_{_{N=4}} &=&
1/2  + 1/8\tau+0.05902778\tau^2+0.02074731\tau^3+
               0.01004792\tau^4+0.00133632\tau^5+
               0.00248012\tau^6,
\nonumber
\\
\nu_{_{N=5}} &=&
1/2 + 2/15\tau+0.05964701\tau^2+0.01877086\tau^3+
               0.00688561\tau^4-0.00041380\tau^5+
               0.00060891\tau^6,
\nonumber
\\
\nu_{_{N=\infty}}&=&
1/2+   1/6\tau+0.03612254\tau^2+0.00709205\tau^3-
               0.00142535\tau^4+0.00103317\tau^5-
               0.00285768\tau^6,
\nonumber
\\
\omega_{_{N=3}}&=&
\tau-          0.39042829\tau^2+0.29428918\tau^{3}-
               0.25565542\tau^4+0.31134025\tau^{5}-
               0.43957722\tau^6,
\nonumber
\\
\omega_{_{N=4}}&=&
\tau-          0.36419753\tau^2+0.24511892\tau^3-
               0.20419925\tau^4+0.21874431\tau^5-
               0.27962773\tau^6,
\nonumber
\\
\omega_{_{N=5}}&=&
\tau-          0.35129140\tau^2+0.21196053\tau^3-
               0.16985912\tau^4+0.17369321\tau^5-
               0.19948859\tau^6,
\nonumber
\\
\omega_{_{N=\infty}}&=&
\tau-          0.42249657\tau^2+0.34513141\tau^3-
               0.32006198\tau^4+0.44947688\tau^5-
               0.67842170\tau^6.
\nonumber
\end{eqnarray}
One can determine the cubic model critical exponents of the new 
universality class on the basis of the expansions (\ref{exp}) in the 
same manner as for the expansion (\ref{nc_tau}) of $N_c$.  However, 
the expansions for the combination $1/\gamma$ and $1/\nu-1$ appear to 
have better convergence properties and all values are obtained on 
their basis.  The Pad\'e-- and Pad\'e--Borel analysis lead to the 
critical exponents values as they are given in the table \ref{tab} in 
the last column (here, we do not show intermediate results similar to 
(\ref{nc_pd})-(\ref{nc_pb})). The error bars for the critical 
exponents given in the Table \ref{tab} were obtained from the maximal 
deviation between the six- and the five-loop results among {\bf all} 
deviations for the parameter $p$ value $0 \leq p \leq 10$.  The error 
bars within the pseudo--$\varepsilon$ expansion are typically much 
smaller than those based on other methods.  The reason is that 
$\beta$-- and $\gamma$--functions contribute in a self-consistent way 
into the pseudo--$\varepsilon$ expansion series for critical 
exponents.

To compare our results we represent in the table \ref{tab} the values 
obtained from the five--loop $\varepsilon$--expansion 
\cite{Kleinert95a} by means of a modified Borel summation 
\cite{Mudrov98} and of the Borel transformation with conformal mapping 
\cite{Carmona99} (the corresponding citation in the table is primed). 
Critical exponents from the four--loop fixed massive $d=3$ scheme with 
an application to the RG functions of Pad\'e-Borel resummation 
\cite{Varnashev99} and the results from the six--loop RG functions 
resummed by Borel transformation with conformal mapping 
\cite{Carmona99} are given in the table. Recently, the modified 
Pad\'e-Borel resummation  \cite{Mudrov98} has been applied 
\cite{Varnashev00} to the six--loop RG functions \cite{Carmona99} of 
the cubic model. These data are also displayed in the table.

The error bars for the values of the Refs. \cite{Carmona99} and 
\cite{Varnashev00} were obtained from the condition of the result 
stability in successive approximation orders.
However, the numerical value of the fixed point 
coordinates were substituted into the expansions for the 
$\gamma$--functions (\ref{gamma_phi}), (\ref{gamma_phi2}). To this end 
the most reliable numerical values of the stable fixed point 
coordinates were substituted into the resummed $\gamma$--functions and 
then an optimal value of the fit parameter in modified Pad\'e-Borel 
resummation \cite{ Varnashev00} and two fit parameters in the 
conformal mapping procedure \cite{Carmona99} were chosen. The 
deviations between five- and six-loop results obtained within the 
resummation procedure with optimal fit parameter(s) value gave the 
error interval.  To complete the list, we show the value of $\omega$ 
for $N=3$ obtained in Ref. \cite{Kl97} on the basis of Borel 
transformation involving knowledge on RG functions large--order 
behaviour. Let us  note that for finite values of $N$ our data for 
$\gamma$ and $\nu$ interpolate the results of the minimal subtraction  
(\cite{Mudrov98},\cite{Carmona99}$^{'}$) and the massive scheme  
(\cite{Varnashev99},\cite{Carmona99}), though the values are closer to 
the last. On the other hand, our method gives smaller values for 
$\omega$ in comparison with other methods. We note as well that 
passing from the four--loop \cite{Varnashev99} to the six--loop 
\cite{Varnashev00} approximation in frames of a massive $d=3$ approach 
shifts the numerical values of critical exponents towards our data.

In the limit $N \rightarrow \infty$ the critical properties of the 
cubic model reconstitute those of the annealed diluted Ising model 
\cite{Aharony73} where Fisher renormalization for critical exponents 
holds \cite{Fisherrenormalization}. In particular, based on the recent 
RG estimates for the critical exponents of the pure Ising model 
$\alpha=0.109 \pm 0.004$, $\nu=0.6304 \pm 0.0013$, $\gamma=1.2397 \pm 
0.0013$ \cite{Guida98}, one obtains the values $\nu=0.708\pm0.005$, 
$\gamma=1.391\pm0.008$ for the Ising model model with  annealed 
disorder.  The last values agree very well with our results of the 
last row of the table \ref{tab}.  Moreover, they are in very good 
agreement with other data of the table.

It is worth to note here that the RG series for the cubic model allow 
to reconstitute the functions which describe the Ising model with the 
other type of randomness. By substitution $N=0$ one reconstitutes the 
weakly diluted quenched Ising model (RIM) \cite{RIM}. In this case, 
however, the pseudo--$\varepsilon$ expansion in $\tau$ degenerates 
into a $\sqrt{\tau}$--expansion for the same reasons as the 
$\varepsilon$--expansion for the RG functions degenerates into a 
$\sqrt{\varepsilon}$--expansion \cite{RIM}. Moreover, our calculation 
show that an expansion in $\sqrt{\tau}$ is numerically useless as this 
was shown for $\sqrt{\varepsilon}$--expansion \cite{Shalaev97, 
Holovatch99a,sqrt}.  This can be regarded as an evidence of the Borel 
non--summability of the RIM RG functions.  Since the asymptotic 
properties of the series still are not proven despite of noticing 
their divergent character \cite{Borelnonsummability,Alvarez99}, the RG 
functions of RIM as series in renormalized couplings used to be 
treated by means of Pad\'e--Borel or Chisholm-Borel resummations (see 
e.g. \cite{Mayer,Pakhnin99,Holovatch99a}).  
The first of the stated methods was 
recently applied to study the five-loop RG functions of RIM 
\cite{Pakhnin99}. The analysis allowed the authors to obtain the 
five--loop estimates for the RIM critical exponents. Extending the 
analysis of Ref. \cite{Pakhnin99} to the six--loop order reveals the 
wide gap between five-- and six--loop fixed point coordinates.  This 
leads to an inconsistency of the six--loop values of critical 
exponents compared with the five--loop results of Ref. 
\cite{Pakhnin99}.  However, the analytical solution of a toy $d=0$ RIM 
showed its free energy to be Borel summable provided that resummation 
is done asymmetrically:  resumming first the series in the coupling $u$ 
and subsequently the series in $v$ \cite{Alvarez99}. The corresponding 
resummation applied to $d=3$ RIM massive scheme RG functions allowed 
precise determination of the critical exponents \cite{Pel}.

\section{Conclusions}
In the present paper we studied the critical properties of a cubic 
model associated with $\phi^4$--terms of spheric and cubic symmetry of 
the LGW Hamiltonian. In particular, we were interested in the 
crossover between $O(N)$--symmetric and cubic behaviour which occurs at 
a certain value $N_c$ of order parameter components number.  Recently, 
five-- \cite{Pakhnin99} and six--loop \cite{Carmona99} order RG 
functions were obtained for the cubic model within massive $d=3$ 
scheme \cite{Parisi}. We applied the pseudo--$\varepsilon$ expansion 
\cite{Nickel} to their analysis. This method is known as a 
standard one for the $O(N)$--symmetric model  analysis and leads to 
the most accurate values of critical exponents \cite{Guida98}. Here, 
to our knowledge, it has been applied to the cubic model for the first 
time \cite{VonFerber}.

The pseudo--$\varepsilon$ expansion for $N_c$ appears to have much 
better convergence properties then the corresponding 
$\varepsilon$--expansion (c.f. Pad\'e--tables (\ref{nc_tau}) and 
(\ref{nc_epsilon})). This provides very good convergence of its 
Pad\'e--analysis (\ref{nc_pd}). The last together with the refined 
Pad\'e--Borel analysis yields the best estimate $N_c=2.862 \pm 0.005$ 
of the paper.  Our conclusion $N_c<3$ means in particular, that all 
ferromagnetic cubic crystals with three easy axis should undergo a 
first order phase transition \cite{Sznajd}.

We obtained the values of cubic model critical exponents in the new 
universality class in pseudo--$\varepsilon$ expansions with the 
results given in table \ref{tab}.  In the $N \rightarrow \infty$ limit 
our data reproduce the critical behaviour of an annealed weakly diluted 
Ising model \cite{Aharony73}.  The $N \rightarrow 0$ limit, 
corresponding to a quenched weakly diluted Ising model \cite{RIM}, 
however does not yield reliable results in pseudo--$\sqrt\varepsilon$
expansion. Within a traditional $d=3$ massive technique
the resummation of the RG functions by means of the convenient 
Pad\'e-Borel analysis reveals a gap between five-- and six--loop fixed 
point coordinates. This leads to an inconsistency of the obtained 
critical exponents values compared to the declared in Ref. 
\cite{Pakhnin99}.  Let us note, however, that recently reliable values 
have been obtained \cite{Pel} by a resummation method which treats the 
couplings of the RIM model asymmetrically \cite{Alvarez99}.

\section*{Acknowledgements}

We acknowledge useful discussions with J\'ozef Sznajd and Maciej 
Dudzi\.nski and thank Konstantin Varnashev for communicating his 
results \cite{Varnashev00} prior to publication.  This work has been 
supported in part by "\"Osterreichische Nationalbank 
Jubil\"aumsfonds'' through the grant No 7694.

\begin{table}
\caption{\label{tab} Our data for the critical exponents of the cubic model (last column)
in comparison with other results. See the text for a full description.}
\begin{center}
\tabcolsep1.6mm
\begin{tabular}{|c|ccccccc|}
%\hline
$N$ && \cite{Mudrov98} & \cite{Carmona99}$^{'}$ & \cite{Varnashev99} &
\cite{Varnashev00} & \cite{Carmona99} & this study \\
\hline
 & $\gamma$ & $1.3746\pm0.0020$  & $1.377(6)$ & $1.3775$ &$1.3850 \pm 0.0050$ &
$1.390(12)$ & $1.387 \pm 0.001$\\
3 & $\nu$    & $0.6997\pm0.0024$  & $0.701(4)$ &$0.6996$ & $0.7040 \pm 0.0040$ &
$0.706(6)$  & $0.705 \pm 0.001$\\
  & $\omega$ & $0.8061$\cite{Kl97}& $0.799(14)$ &$0.7786$ & $0.7833\pm 0.0054$ &
$0.781(4)$  & $0.777 \pm 0.009$\\
\hline
  & $\gamma$ & $1.4208\pm0.0030$  & $1.419(6)$ & $1.4028$ & $1.4074\pm 0.0030$ &
$1.405(10)$ & $1.416 \pm 0.004$\\
4 & $\nu$    & $0.7225\pm0.0022$  & $0.723(4)$ & $0.7131$ & $0.7150\pm 0.0050$ &
$0.714(8)$  & $0.719 \pm 0.002$\\
  & $\omega$ &    ---             & $0.790(8)$ &---     & $0.7887\pm 0.0090$ &
$0.781(44)$  & $0.777 \pm 0.002$\\
\hline
  & $\gamma$ & $1.4305\pm0.0040$  & ---        &$1.4076$ & --- &
---         & $1.417 \pm 0.006$\\
5 & $\nu$    & $0.7290\pm0.0016$  & ---        &$0.7154$ & --- &
 ---         & $0.720 \pm 0.004$\\
  & $\omega$ &    ---             & ---        & ---     &  --- &
---        & $0.773 \pm 0.003$\\
\hline
  & $\gamma$ & $1.4322\pm0.0040$  & ---        & $1.4082$ & --- &
---         & $1.417 \pm 0.009$\\
6 & $\nu$    & $0.7301\pm0.0016$  & ---        & $0.7157$ & --- &
 ---         & $0.718 \pm 0.003$\\
  & $\omega$ &    ---             & ---        & ---     &  --- &
---        & $0.771 \pm 0.005$\\
\hline
  & $\gamma$ & ---               & $1.422(6)$ &$1.4074$ & $1.4068\pm 0.0030$ &
$1.404(10)$         & $1.407 \pm 0.008$\\
8 & $\nu$    & ---                & $0.723(2)$ &$0.7153$ & $0.7143 \pm 0.0035$ &
$0.712(6)$         & $0.715 \pm 0.003$\\
  & $\omega$ &    ---             & $0.786(6)$ & ---     &  $0.7955\pm 0.0150$ &
$0.775(88)$        & $0.770 \pm 0.009$\\
\hline
  & $\gamma$ & $1.3993\pm0.0020$  & $1.399(8)$ &---      & $1.3962\pm 0.0040$ &
$1.396(14)$ & $1.395 \pm 0.006$\\
$\infty$
  & $\nu$    & $0.7108\pm0.0010$  & $0.711(2)$ &---      &  $0.7094\pm 0.0030$ &
$0.708(8)$ & $0.708 \pm 0.001$\\
  & $\omega$ &    ---             & $0.802(18)$ & ---     & $0.7986\pm 0.0200$ &
$0.790(18)$& $0.775 \pm 0.020$\\
\end{tabular}
\end{center}
\end{table}


\begin{thebibliography}{99}
\bibitem{dependencefree}
Though a set of resulting RG functions are renormalization dependent, measurable
quantities are universal and do not depend on the renormalization scheme.
\bibitem{normalization}
Here, a convenient normalization is used where coefficients of the
one--loop contribution at the coupling $u$ for $\beta_u$ and $v$ for
$\beta_v$ equal $1$.
\end{thebibliography}
\end{document}